\documentclass[preprint,12pt]{elsarticle}

\usepackage{amssymb}
\usepackage{amsmath}
\usepackage{amsthm}
\usepackage{xspace}
\usepackage{lineno}
\usepackage{hyperref}
\usepackage{cleveref}

\theoremstyle{plain}
\newtheorem{theorem}{Theorem}              
\newtheorem{lemma}[theorem]{Lemma}  
\newtheorem{observation}[theorem]{Observation}    
\newtheorem{corollary}[theorem]{Corollary}

\theoremstyle{definition}
\newtheorem{definition}[theorem]{Definition}

\crefname{theorem}{Theorem}{Theorems}
\crefname{lemma}{Lemma}{Lemmas}
\crefname{corollary}{Corollary}{Corollaries}
\crefname{proposition}{Proposition}{Propositions}
\crefname{definition}{Definition}{Definitions}

\newcommand{\dom}{\textsc{Dominating Set}\xspace}

\newcommand{\gmegset}{\textsc{GMEG-Set}\xspace}
\newcommand{\megset}{\textsc{MEG-Set}\xspace}
\newcommand{\p}{\ensuremath{\mathsf{P}}\xspace}
\newcommand{\np}{\ensuremath{\mathsf{NP}}\xspace}
\newcommand{\apx}{\ensuremath{\mathsf{APX}}\xspace}

\begin{document}

\begin{frontmatter}

\title{On the (In)Approximability of the Monitoring Edge-Geodetic Set Problem}

\author[inst1]{Davide Bil\`o\corref{cor1}}
\ead{davide.bilo@univaq.it}
\ead[url]{https://orcid.org/0000-0003-3169-4300}

\author[inst1]{Giordano Colli\corref{cor1}}
\ead{giordano.colli@graduate.univaq.it}
\ead[url]{https://orcid.org/0009-0000-8361-2333}

\author[inst1]{Luca Forlizzi}
\ead{luca.forlizzi@univaq.it}
\ead[url]{https://orcid.org/0000-0002-3923-7668}

\author[inst1]{Stefano Leucci}
\ead{stefano.leucci@univaq.it}
\ead[url]{https://orcid.org/0000-0002-8848-7006}

\cortext[cor1]{Corresponding authors}

\affiliation[inst1]{organization={Department of Information Engineering, Computer Science, and Mathematics, University of L'Aquila},
            city={L'Aquila},
            country={Italy}}

\begin{abstract}
    The minimum \emph{Monitoring Edge-Geodetic Set} problem asks for a minimum vertex subset $M$ such that for every edge $e$ of the input graph $G$, there exist a two vertices $u,v \in M$, that monitor $e$, i.e. every shortest path between $u$ and $v$ in $G$ contains $e$.

We prove that this problem cannot be approximated within an $\Omega(\log n)$ ratio in polynomial time unless $\np \subseteq \mathsf{ZTIME}(n^{O(\log \log n)})$. Additionally, we establish hardness and conditional running-time lower bounds for exact algorithms on $2$-degenerate, $3$-degenerate, and $1$-apex $2$-degenerate graphs.

 Moreover, we design an approximation algorithm for hereditary graph classes with efficiently computable, truly sublinear-size balanced separators. This achieves an $O(n^{\frac{1}{4}} \sqrt{\log n})$ approximation ratio for $K_h$-minor-free graphs ($h=O(1)$), bounded genus graphs, and $k$-apex graphs ($k=O(n^{\frac{1}{4}})$). For graphs with bounded treewidth, the algorithm yields an $O(\log^{3/2} n)$ approximation ratio.
\end{abstract}


\end{frontmatter}


\section{Introduction}
    Consider a communication network that might suffer link failures. Such a network can be modeled as an undirected and connected graph $G$, where vertices represent hosts and edges correspond to links between hosts. 
    In order to quickly detect failure events, the designer wishes to equip the hosts of the network with additional \emph{probes}: probes can communicate with one another and they can monitor the distance between their corresponding nodes in the network (e.g., via a traceroute mechanism).
    Whenever the distance between a pair of vertices increases, this indicates that some communication link is inoperative, and an alarm can be raised. Observe that, in the above scenario, it may be possible for two probes to be connected by two (or more) edge-disjoint shortest paths, in which case they will not be able to detect an edge failure.
    
    This motivates the following problem: given a network $G$, find the smallest possible set $M$ of vertices to equip with probes while ensuring that each edge $e$ of $G$ has at least one pair of vertices $u,v \in M$ for which all shortest paths between $u$ and $v$ in $G$ traverse $e$.
    The above problem is known as the minimum \emph{Monitoring Edge-Geodetic Set} (\megset) problem and was introduced in \cite{Monitoring_edge-geodetic_sets_in_graphs}, where the authors focus on providing upper and lower bounds on the size of minimum \megset{}s for both general graphs and special graph classes (trees, cycles, unicyclic graphs, complete graphs, grids, hypercubes, \dots).
    
    Further bounds on the size of \megset{}s have been given for the Cartesian and strong products of two graphs \cite{haslegrave}, for other graph products including join and direct products \cite{XuYBZS24}, as a function of the graph's girth and chromatic number \cite{FoucaudMMSST24}, for ladder, butterfly, circulant and Benes networks, for convex polytopes \cite{TanLWL23}, and for radix triangular mesh networks and Sierpiński graphs \cite{MaJYL24}.
    Moreover, the minimum \megset problem was proven to be \np-hard on general graphs \cite{haslegrave}, \np-hard on $3$-degenerate $2$-apex graphs \cite{FoucaudMMSST24_Caldam24, foucaud2024algorithmscomplexitymonitoringedgegeodetic}, and \apx-hard on $4$-degenerate graphs \cite{foucaud2024algorithmscomplexitymonitoringedgegeodetic}. If the Exponential Time Hypothesis \cite{ImpagliazzoPZ01} holds, then the problem cannot be solved in subexponential time on $3$-degenerate graphs \cite{foucaud2024algorithmscomplexitymonitoringedgegeodetic}. In the same paper, the authors also provide an exact polynomial-time algorithm for interval graphs and two FPT algorithms: one for general graphs parameterized by the sum of cliquewidth and diameter, and the other for chordal graphs parameterized by treewidth.
    Polynomial-time algorithms for computing optimal \megset{}s are also known for distance-hereditary graphs, $P_4$-sparse graphs, bipartite permutation graphs, and strongly chordal graphs \cite{foucaud2025characterizingoptimalmonitoringedgegeodetic}.

    As far as approximation algorithms are concerned, it has been observed \cite{Colli2023, BiloCF024, foucaud2024algorithmscomplexitymonitoringedgegeodetic} that a simple reduction to the \textsc{Set Cover} problem yields a polynomial-time algorithm returning solutions of size $O(k^2 \log n)$, where $k$ is the size of a \megset of minimum cardinality,
    thus achieving an approximation ratio of $O(\min\{ k \log n, n/k \}) = O(\sqrt{n \log n})$.\footnote{To the best of our knowledge, this algorithm was first given and analyzed in a Bachelor's thesis \cite{Colli2023}, %
    and the result was subsequently claimed in a short communication based thereon \cite{BiloCF024}. The same algorithm, and its analysis, also appear in a later unpublished work \cite{foucaud2024algorithmscomplexitymonitoringedgegeodetic} by a different group of authors.}

    \subsection{Our results}

    In this paper, we investigate the approximability and inapproximability of the minimum \megset problem.

    We prove that, unless $\np \subseteq \mathsf{ZTIME}(n^{O(\log \log n)})$, all polynomial-time (possibly randomized)  approximation algorithms for \megset must have an approximation factor of $\Omega(\log n)$. To the best of our knowledge, these are the first non-constant inapproximability results for the minimum \megset problem. Moreover, if $\p \neq \np$, \megset is not approximable within a constant factor better than $10\sqrt{5}-21 \approx 1.36$ on graphs with degeneracy at most $3$, which can be improved to $2$ if the unique games conjecture holds \cite{Khot02a}. As a comparison, \cite{foucaud2024algorithmscomplexitymonitoringedgegeodetic} only proves \apx-Hardness for $3$-degenerate graphs.

    We also show that the problem remains \np-hard even on $2$-degenerate $1$-apex graphs (while the reduction of \cite{FoucaudMMSST24_Caldam24,foucaud2024algorithmscomplexitymonitoringedgegeodetic} uses $3$-degenerate $2$-apex graphs), and that it cannot be solved in time $2^{o(\sqrt{n})}$ even on sparse graphs unless the Exponential Time Hypothesis (ETH) fails. 
    If the graph is $2$-degenerate, but we impose no constraints on its apex number, then \megset cannot be solved in time $2^{o(n)}$, assuming the ETH. This improves a similar result of  \cite{foucaud2024algorithmscomplexitymonitoringedgegeodetic} for $3$-degenerate graphs.

    We also extend the aforementioned $O(\sqrt{n \log n})$-approximation algorithm for minimum \megset to a generalized version of the problem in which edges have non-negative \emph{lengths}, vertices have binary \emph{costs}, and a minimum-cost subset of vertices that monitors \emph{a given subset} of edges needs to be selected. We use this result as a building block to design a more involved approximation algorithm for minimum \megset that relies on recursively computing vertex-separators and provides good approximations for graph families that admit sparse balanced vertex-separators.
    More precisely, we achieve an approximation ratio of $O(n^{\frac{1}{4}} \sqrt{\log n})$ on $K_h$-minor-free graphs with $h=O(1)$ (which include planar graphs), on graphs with bounded genus, and on $k$-apex graphs when $k=O(n^{\frac{1}{4}})$. 
    We also obtain an approximation ratio of $O(\log^{3/2} n)$ on graphs with bounded treewidth.
    It is an interesting open problem to settle the complexity status of the problem for planar graphs.

\subsection{Organization of the paper}

We start by giving some preliminary definitions and notation in \Cref{sec:prelim}. 
Our \np-hardness results, along with our lower bounds based on ETH are given in \Cref{sec:apex1-hardness}. Our inapproximability results are discussed in \Cref{inapproximability_section}.
Finally, \Cref{sec:apx_algs} is devoted to our approximation algorithms. 
 
\section{Preliminaries and notation}
\label{sec:prelim}
Throughout the paper, we consider simple, connected, and undirected graphs unless otherwise specified. 
Given a graph $G = (V,E)$, we may use $V(G)$ or $E(G)$ to refer to the vertex-set $V$ or the edge-set $E$, respectively. We denote the open neighborhood of a vertex $v \in V(G)$ in a graph $G$ as $N_G(v) = \{u \in V(G): \{u,v\} \in E(G)\}$; in the same way, the closed neighborhood of a vertex $v \in V(G)$ in $G$ is defined as $N_G[v] = N_G(v) \cup \{v\}$. When the graph is clear from the context, we refer to the open (resp.\ closed) neighborhood of a vertex $v \in V(G)$ as $N(v)$ (resp.\ $N[v]$). We also use $n$ to refer to $|V(G)|$. For a set $C\subseteq V$, we define $N(C)=\bigcup_{v \in C}N(v)$ and $N[C]=\bigcup_{v \in C}N[v]$.

We say that an edge $(u,v) \in E(G)$ is \emph{monitored} by a pair of distinct vertices $x,y \in V(G)$ if $(u,v)$ lies on \emph{all} the \emph{shortest} paths between $u$ and $v$. Similarly, a set of edges $E'$ is monitored by $M \subseteq V$ if each edge $e \in E'$ is monitored by at least one pair of vertices in $M$. A \emph{monitoring edge-geodetic set} \megset of $G = (V,E)$ is a subset of vertices that monitors~$E$.

A walk $\pi$ between two vertices $u$ and $v$ in an edge-weighted graph $G$ with weight function $w$ is an ordered list of alternating (and not necessarily distinct) vertices and edges $\langle v_0, e_1, v_1, e_2, v_2, \dots, e_{k}, v_k\rangle$ with $v_0 = u$, $v_k = v$, and such that $e_i = (v_{i-1}, v_i)$ for all $i=1, \dots, k$. The weight $w(\pi)$ of $\pi$ is $\sum_{i=1}^{k} w(e_i)$.

We conclude this section by stating two well-known structural properties of \megset{}s.

    \begin{lemma}[\cite{Colli2023,Monitoring_edge-geodetic_sets_in_graphs}]
        \label[lemma]{lemma:leaf}
        All vertices of degree $1$ in $G$ belong to all \megset{}s of $G$.
    \end{lemma}
    
    \begin{lemma}[\cite{Colli2023,FoucaudMMSST24}]
        \label[lemma]{lemma:leaf_neighbor}
        Let $u$ be a vertex of degree $1$ in $G$ and let $v$ be its sole neighbor. If $|V(G)| \ge 3$ and $M$ is a \megset{} of $G$, then $M \setminus \{v\}$ is a \megset{} of $G$.
    \end{lemma}

\section{Negative results on graphs with degeneracy \texorpdfstring{$2$}{2}}
\label{sec:apex1-hardness}
    We start by providing a reduction from the minimum \dom problem on graphs with girth at least seven (which is known to be \np-Hard even for planar graphs \cite[Problem GT2]{GareyJ79}) to the minimum \megset problem.

    Let $G = (V,E)$ be an input graph with $|V(G)| \ge 3$ and with girth $g(G) \ge 7$. 
    We construct an instance of \megset, namely $H$, starting from $G$ and augmenting it as follows: for each vertex $v \in V(G)$, we add two vertices $v'$ and $v''$ to $G$, then connect $v$ to $v'$ and $v'$ to $v''$.
    We denote the set containing all the vertices $v'$ as $L'$ and the set containing all the vertices $v''$ as $L''$. Notice that: 
    \begin{enumerate}
      \item[i)] Each $v'' \in L''$ belongs to every \megset of $G$ due to Lemma \ref{lemma:leaf};
        \item[ii)] Each $v' \in L'$ does not belong to any optimal \megset of $G$ due to Lemma \ref{lemma:leaf_neighbor}.
    \end{enumerate}
    The basic idea is to monitor each edge that is incident to a vertex $v \in V(G)$ that belongs to a \emph{selected} \megset in $H$. We denote the set of edges $(v,v')$ as $E_{L'}$ and the set of edges $(v',v'')$ as $E_{L''}$. Note that at this point of the construction, all edges are monitored by at least one pair of leaves in $L''$. To avoid this, we shortcut all the shortest paths from one vertex in $L''$ to another by inserting two vertices $v_{*}$ and $v_{*}'$ and connecting them together. Then, we connect $v_{*}$ to all vertices $v' \in L'$. We denote the set containing edges $(v_*,v')$ with $v' \in L'$ as $E_*$. The vertex $v_{*}$ is used to create shortcuts between each shortest path from a vertex in $L' \cup L''$ to a vertex in $L' \cup L''$. Notice that, since $v_{*}'$ is a leaf, it belongs to every \megset of $G$ due to \Cref{lemma:leaf}. The vertex $v_{*}'$ is used to monitor the edges in $E_*$. 
     Formally, let $H = (V(G) \cup L' \cup L'' \cup \{v_*,v_{*}'\},E(G) \cup E_{L'} \cup E_{L''} \cup E_* \cup \{ \;  \{v_*,v_{*}'\} \; \}) $, where:
    \begin{itemize}
        \item $L' = \{v': v \in V(G)\}$;
        \item $L'' = \{v'': v \in V(G)\}$;
        \item $E_{L'} = \{\{v,v'\}: v \in V(G)\}$;
        \item $E_{L''} = \{\{v'',v'\}:  v \in V(G)\}$;
        \item $E_{*} = \{\{v_*,v'\}:  v \in V(G)\}$.
    \end{itemize}

\begin{figure}[t]
    \centering
    \includegraphics[scale=0.85]{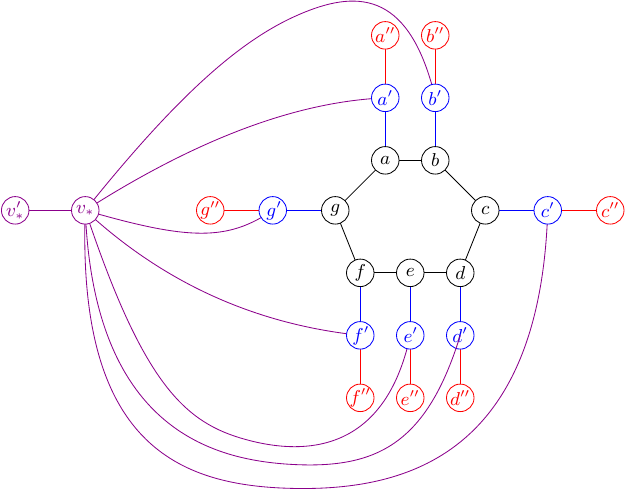}
    \caption{An example of the construction of the graph $H$ from the input graph $G = (\{a,b,c,d,e,f,g\},\{ \, \{a,b\}, \{b,c\}, \{c,d\}, \{d,e\}, \{e,f\}, \{f,g\} \{g,a\} \, \})$, which is shown in black. }
    \label{constructionPicture}
\end{figure}

We now prove two lemmas.
 The first lemma shows that if there is a dominating set of the graph $G$ having size at most $k$, there exists a \megset of $H$ having size $k + |V(G)| + 1$. Conversely, the second lemma shows that if there is a \megset of the graph $H$ having size at most $k + |V(G)| + 1$, there exists a dominating set of $G$ having size at most $k$.
\begin{lemma}
    \label[lemma]{dominatingSetToMeg}
    Let $D$ be a dominating set of $G$ of size at most $k$. Then $M = D \cup L'' \cup \{v_*'\}$ is a \megset of $H$ and $|M| \le k + |V(G)| + 1$.
\end{lemma}
\begin{proof}
    We prove that each edge $e \in E(H)$ is monitored by $M$. We distinguish four cases:
    \begin{itemize}
        \item Case 1: $e \in E(G)$. Let $e = (u,v)$. Since $D$ is a dominating set of $G$, there are two subcases:
        \begin{itemize}
            \item At least one of $u$ and $v$ belongs to $D$, therefore we assume w.l.o.g.\ that $u \in D$. Each path from $v''$ to $u$ must either pass through the vertex $v$ or pass through the vertex $v_*$. The unique shortest path including $v$ is trivially $\langle (v'',v'), (v',v), (v,u) \rangle$ and has length $3$. Each path passing through $v_*$ must contain at least four edges. Therefore, $(u,v)$ is monitored by $\{v'',u\}$, where both $v''$ and $u$ are in $M$ since $v'' \in L''$ and $u \in D$.
            \item Neither $u$ nor $v$ is in $D$. First, we observe that $u$ and $v$ are dominated by two distinct vertices in $D$; otherwise, there would be an induced triangle in $G$, which is not allowed due to $g(G) \ge 7$. Let these two distinct vertices be $\{x,y\} \subseteq D$. They have distance $d(x,y)$ equal to $3$ since, there is no alternative shortest path between $x$ and $y$ in $H$ by $g(G) \ge 7$.
            Thus the pair $\{x,y\}$ monitors the $(u,v)$ edge. 
        \end{itemize}
        \item Case 2: $e \in E_{L''} \cup E_{*}$. In this case, there is \textit{exactly} one endpoint $u'$ of $e$ that belongs to $L'$. It follows that either $e = (u'', u')$ with $u'' \in L''$ or $e = (v_*, u')$. Notice that there is a unique shortest path from $v_*'$ to $u''$ including the edges $(v_*',v_*),(v_*,u'),(u',u'')$. This implies that the edge $e$ is monitored by $\{v'_*, u''\}$, where both $v'_*$ and $u'' \in L''$ are in $M$.
        
        \item Case 3: $e \in E_{L'}$. Let $e = (u,u')$ where $u$ is the unique endpoint of $e$ in $V(G)$. Since $D$ dominates all the vertices in $V(G)$,
        either $u \in D$ or $u \not\in D$ and there exists some vertex $x \in D \cap N[u]$. There is a single shortest path from $u''$ to $x$, and it traverses the edge $(u,u')$.  
        
        \item Case 4: $e = (v'_*, v_*)$. All shortest paths from $v'_*$ to any other vertex in $H$ must traverse $e$, and hence $e$ is monitored, e.g., by all pairs  $\{v'_*, u\}$ with $u \in D$. Note that there exists at least one vertex $u \neq v'_*$ in $M$ since $|M| \ge 2$.   
    \end{itemize}
    
The cardinality $|M|$ is equal to the sum of the cardinality of $M \setminus V(G)$ and the cardinality of $M \cap V(G)$, with $(M \setminus V(G)) \cap (M \cap V(G)) = \emptyset$. Since $|M \cap V(G)|$ is equal to the size of the dominating set of $G$, namely $k$, and $M \setminus V(G) = L'' \cup \{v_*'\}$, we can conclude that:
\begin{equation*}
|M| = |M \cap V(G)| + |M \setminus V(G)| = k + |L''| + |\{v_*'\}| = k + |V(G)| + 1. \qedhere
\end{equation*}
\end{proof}

\begin{lemma}
    \label[lemma]{megToDominatingSet}
    Let $M$ be a \megset of $H$ with size at most $k + |V(G)| + 1$ (for a suitable value of $k$). Then, $M \cap V(G)$ is a dominating set of $G$ with size at most $k$.
\end{lemma}
\begin{proof}
    Let $M' = M \setminus (L' \cup \{v_*\})$ be the set obtained by removing the vertex $v_*$ and all the vertices of $L'$ from $M$. Due to Lemma \ref{lemma:leaf_neighbor}, $M'$ is a \megset of $G$. Moreover, since $(L' \cup \{v_*\}) \cap V(G) = \emptyset$, we can observe that $M' \cap V(G) = M \cap V(G)$.
    Let $D = M' \setminus (L'' \cup \{v_*'\})$ be the set of vertices in the \megset{} $M'$ that are in $V(G)$. 
    We prove that if $v \in V(G)$, then either $v \in D$ or $N(v) \cap D \neq \emptyset$. 
    Towards a contradiction, suppose that $v \not\in D$ and no vertices in $N(v)$ are in $D$. Since $M'$ is a \megset{} of $H$, it monitors every edge in $E(G) \subseteq E(H)$. Consider an edge $(u,v)$ with $v \in N(u)$. $(u,v)$ must be monitored by a pair of vertices $x,y$ not in $\{v\} \cup N(v)$. It follows that the distance between $x$ and $y$ is at least four, but there is a path between $x$ and $y$ of length four that only uses edges in $E(H) \setminus E(G)$ passing through $v_*$; thus, the edge $(u,v)$ cannot be monitored by the pair $\{x,y\}$, which is a contradiction.
    The above discussion shows that $M' \cap V(G)$ is a dominating set of $G$. The size of such a dominating set is $|M' \cap V(G)| \le |M \cap V(G)| = |M| - |M \setminus V(G)| \le |M| - (|L''| + 1) \le |M| - (|V(G)| + 1) \le k$, where we used $(L'' \cup \{v'_*\}) \subseteq M \setminus V(G)$, as ensured by Lemma \ref{lemma:leaf}.
\end{proof}

Observe that deleting vertex $v_*$ results in a graph having the same apex number of $G$ (see Figure~\ref{constructionPicture}).
Moreover, one can first delete all vertices in $L'' \cup \{v'_*\}$ (having degree $1$), then all vertices in $L'$ (which now have degree $2$), and then follow any elimination order for $G$. Then, we have the following:
\begin{observation}
    \label{obs:apex}
   Given a graph $G$ with apex number $a$ and degeneracy $d$, the apex number and the degeneracy of the graph obtained by applying the above reduction to $G$ are at most $a+1$ and $\max\{2,d\}$, respectively.
\end{observation}

\begin{theorem}
    The \megset decision problem is \np-complete on $1$-apex $2$-degenerate graphs.
\end{theorem}
\begin{proof}
    It is known that the \textsc{Dominating Set} problem is \np-complete on planar graphs \cite[Problem GT2]{GareyJ79}. 
    Moreover, any planar graph $G$ with $m$ edges can be transformed in polynomial time into a planar graph $G'$ with degeneracy $2$ such that $G$ has a dominating set of size at most $k$ if and only if $G'$ has a dominating set of size at most $k + m$. Indeed, one can obtain $G'$ by replacing each edge $(u,v)$ of $G$ with a path between $u$ and $v$ of length $4$ passing through $3$ internal split vertices. Observe that the degeneracy of $G'$ is at most $2$ since deleting all split vertices (of degree $2$) results in the empty graph.
    This shows that \textsc{Dominating Set} is \np-complete on planar graphs with degeneracy at most $2$ and girth at least $7$.

    Given such an instance $G'$ of \textsc{Dominating Set}, it is possible to construct in polynomial time a graph $H$ using the reduction discussed above.
    From Lemmas \ref{dominatingSetToMeg} and \ref{megToDominatingSet}, there exists a dominating set $D$ of $G'$ such that $|D| \le k$ if and only if there exists a \megset $M$ of $H$ such that $|M| \le k + |V(G')| + 1$. 
    Thanks to Observation~\ref{obs:apex}, we have that the \megset decision problem is \np-Hard on $1$-apex $2$-degenerate graphs.
\end{proof}

Assuming the Exponential Time Hypothesis (ETH) there is no algorithm that decides whether a $3$-CNF-SAT formula on $n'$ variables is satisfiable in time $2^{o(n')}$, and this holds even for formulas in which the number of clauses is $O(n')$ \cite{ImpagliazzoPZ01,LokshtanovMS11}.
By chaining the textbook reduction from \textsc{$3$-CNF-SAT} to \textsc{Vertex Cover} \cite[Problem GT1, \S 3.1.3]{GareyJ79}, 
with the folklore reduction\footnote{The \textsc{Dominating Set} instance is obtained by augmenting the \textsc{Vertex Cover} instance via the addition of a new vertex $x_{u,v}$ for each original edge $(u,v)$, along with the two new edges $(x_{u,v}, u)$ and $(x_{u,v}, v)$.} from \textsc{Vertex Cover} to \textsc{Dominating Set}, and by observing that such reductions produce graphs with a constant-factor blow-up in the number $n = \Theta(n')$ of vertices, and $m = O(n)$ edges, we have that \textsc{Dominating Set} on sparse graphs cannot be solved in time $2^{o(n)}$, unless the ETH fails.
By replacing each edge with a path of length $4$, as before, we can consider instances of \textsc{Dominating Set} with $\Theta(n)$ vertices and edges, girth larger than $7$, and degeneracy at most $2$.
Since our reduction also has a linear blow-up in the number of vertices and edges, we have the following:

\begin{theorem}
    The \megset decision problem on graphs with $n$ vertices, $O(n)$ edges, and degeneracy at most $2$ cannot be solved in time $2^{o(n)}$, unless the Exponential Time Hypothesis fails.
\end{theorem}

Moreover, a $3$-CNF-SAT formula can be made planar with a quadratic blow-up in the number of clauses and variables \cite{Lichtenstein82}, while the above reductions from \textsc{$3$-CNF-SAT} to \textsc{Vertex Cover} and from \textsc{Vertex Cover} to \textsc{Dominating Set} preserve planarity. Since our reduction increases the apex number of the input graph by at most $1$, we have the following:
\begin{theorem}
    The \megset decision problem on $1$-apex graphs with $n$ vertices, $O(n)$ edges, and degeneracy at most $2$ cannot be solved in time $2^{o(\sqrt{n})}$, unless the Exponential Time Hypothesis fails.
\end{theorem}

\section{Inapproximability of \megset}
\label{inapproximability_section}
     In this section we provide inapproximability results for \megset by reducing from the \textsc{Set Cover with intersection $1$} problem. A \textsc{Set Cover with intersection $1$} instance $\mathcal{I}=\langle X, \mathcal{S}\rangle$ is described as a set of $\eta$ items $X = \{x_1, \dots, x_\eta\}$ and a collection $\mathcal{S} = \{ S_1, \dots, S_h\}$ of $h \ge 2$ distinct subsets of $X$, such that $|S_i \cap S_j| \le 1$ for any $1 \le i < j \le h$.
    The goal is that of computing a collection $\mathcal{S}^* \subseteq \mathcal{S}$ of minimum size such that $\cup_{S_i \in \mathcal{S}^*} S_i = X$.\footnote{We assume w.l.o.g.\ that $\cup_{i=1}^h S_i = X$, i.e., that a solution exists.} Observe that $h \le \eta^2$ since, for each item $x_i$, there can be at most $\eta$ distinct subsets $S_j \in \mathcal{S}$ with $x_i \in S_j$.
    It is known that, unless $\mathsf{NP} \subseteq \mathsf{ZTIME}(\eta^{O(\log \log \eta)})$, all polynomial-time randomized approximation algorithms for \textsc{Set Cover with intersection $1$} must have an approximation ratio of $\Omega(\log \eta)$ \cite{KumarAR00}.

    Given an instance $\mathcal{I} = \langle X, \mathcal{S} \rangle$ of \textsc{Set Cover with intersection $1$}, we build an associated bipartite graph $H$ whose vertex set $V(H)$ is $X \cup \mathcal{S}$ and such that $H$ contains the edge $(x_i, S_j)$ if and only if $x_i \in S_j$. We define $N=h+\eta$. %
    
\begin{figure}
        \centering
        \includegraphics[scale=.9]{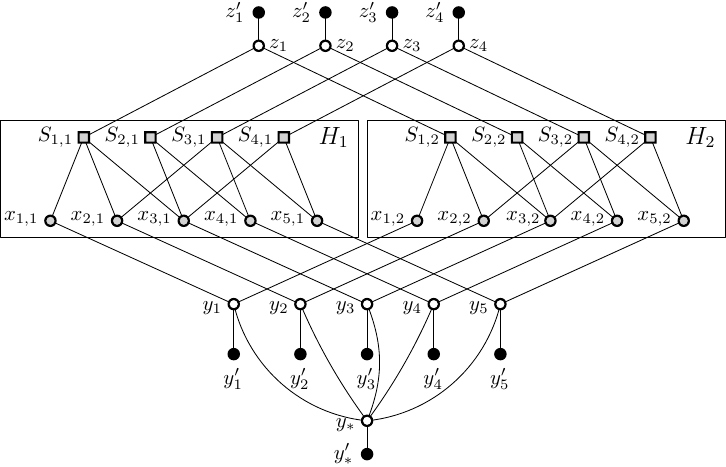}
        \caption{The graph $G$ obtained by applying our reduction with $k=2$ to the \textsc{Set Cover} instance $\mathcal{I} = \langle X, \mathcal{S}\rangle$ with $\eta=5$, $h=4$, $S_1 = \{x_1, x_2, x_3\}$, $S_2 = \{x_3, x_4\}$, $S_3 = \{x_2, x_4, x_5\}$,  and $S_4 = \{x_3, x_5\}$. }
        \label{fig:reduction2}
    \end{figure}

    Let $k$ be an integer parameter, whose exact value will be chosen later, that satisfies $2 \le k = O(\text{poly}(N))$. We construct a graph $G$ that contains $k$ copies $H_1,\ldots,H_k$ of $H$ as induced subgraphs. In the following, for any $\ell=1,\ldots,k$, we denote by $x_{i,\ell}$ and $S_{j,\ell}$ the vertices of $H_\ell$ corresponding to the vertices $x_i$ and $S_j$ of $H$, respectively. More precisely, we build $G$ by starting with a graph that contains exactly the $k$ copies $H_1,\ldots,H_k$ of $H$ and augmenting it as follows (see \Cref{fig:reduction2}):
    \begin{itemize}
        \item For each item $x_i \in X$, we add two new vertices $y_i$, $y'_i$ along with the edge $(y_i, y'_i)$ and all the edges in $\{(x_{i,\ell}, y_i) \mid \ell=1,\ldots, k\}$;
        \item We add two new vertices $y_*, y'_*$ along with edge $(y_*, y'_*)$ and all the edges  $(y_i, y_*)$ for $i = 1, \dots, \eta$;
        \item For each set $S_j \in \mathcal{S}$, we add two new vertices $z_j$, $z'_j$ along with the edge $(z_j, z'_j)$, and all the edges in $\{(S_{j,\ell}, z_j) \mid \ell=1,\ldots, k\}$.
    \end{itemize}
    Observe that the number $n$ of vertices of $G$ satisfies $n= 2h + 2\eta + k(\eta + h) + 2 = (k+2)(\eta + h) + 2 = (k+2)N + 2$.
    
    Let $Y = \{y_i \mid  i=1, \dots, \eta \}$, $Y' = \{y'_i \mid  i=1, \dots, \eta \}$, $Z = \{z_j \mid  j=1, \dots, h \}$, and
    $Z' = \{z_j' \mid  j=1, \dots, h \}$.
    Moreover, define $L$ as the set of all vertices of degree $1$ in $G$, i.e., $L = Y' \cup Z' \cup \{y_*'\}$. By \Cref{lemma:leaf}, the vertices in $L$ belong to all \megset{}s of $G$. 
    
    \begin{lemma}
        \label[lemma]{lemma:edges_in_G_minus_H_monitored}
        $L$ monitors all edges having both endvertices in $Y \cup Y' \cup Z \cup Z' \cup \{ y_*, y'_* \}$.
    \end{lemma}
    \begin{proof}
        Observe that the unique edge incident to a leaf $v \in L$ lies on every shortest path between $v$ and any other vertex in $V$.
        Since $|L| \ge 2$ by construction, all edges incident to some vertex in $L$ are monitored by $L$.
        The only remaining edges incident to some vertex in $L$ are between $y_*$ and some vertex $y_i \in Y$. Let $(y_i, y_*)$ be such an edge. Since the only shortest path between $y'_i$ and $y'_*$ in $G$ is $\langle y'_i, y_i, y_{*}, y'_{*} \rangle$, the pair $\{ y'_*, y'_{i'} \}$ monitors $(y_i, y_*)$.
    \end{proof}
    
    \begin{lemma}\label[lemma]{lemma:set_cover_induces_meg_set}
        Let $\mathcal{S}_1,\ldots,\mathcal{S}_k$ be $k$ (not necessarily distinct) set covers of $\mathcal{I}$. The set $M = L \cup \{S_{j,\ell} \mid S_j \in \mathcal{S}_\ell, 1 \leq \ell \leq k\}$ is a \megset{} of $G$. 
    \end{lemma}
    \begin{proof}
        Since $L \subseteq M$, by \Cref{lemma:edges_in_G_minus_H_monitored}, we only need to argue about edges with at least one endvertex in some $H_\ell$, with $1\leq \ell\leq k$.
        Let $S_j \in \mathcal{S}_\ell$, and consider any $x_i \in S_j$. Edge $(S_{j,\ell},z_j)$ is monitored by $\{z'_j, S_{j,\ell}\}$. Edges $(S_{j,\ell}, x_{i,\ell})$ and $(x_{i,\ell}, y_i)$ are monitored by $\{S_{j,\ell}, y'_i\}$.
    
        The only remaining edges with at least one endvertex in $H_\ell$ are those incident to vertices $S_{j,\ell}$ with $S_j \in \mathcal{S} \setminus \mathcal{S}_\ell$. Consider any such $S_j$, let $x_i$ be an item in $S_j$, and let $S_{k'} \in \mathcal{S}_\ell$ be any set such that $x_i \in S_{k'}$ (notice that both $x_i$ and $S_{k'}$ exist since sets are non-empty and $\mathcal{S}_\ell$ is a set cover).
        Our choice of set cover instance guarantees that $S_{k'} \cap S_j = \{x_i\}$, hence the sole shortest path between $S_{k',\ell}$ and $z'_j$ is $\langle S_{k', \ell}, x_{i,\ell}, S_{j,\ell}, z_j, z'_j \rangle$, which monitors both the edge $(S_{j,\ell}, x_{i,\ell})$ and the edge $(S_{j,\ell}, z_j)$.
    \end{proof}

    \begin{lemma}\label[lemma]{lemma:meg_set_monitors_set_cover}
        Let $M$ be a \megset{} of $G$. For every $i=1,\ldots, \eta$ and every $\ell=1,\ldots,k$, at least one of the following conditions is true: (i) $x_{i,\ell} \in M$; or (ii) there exists an index $j$ such that $x_i \in S_j$ and the set $M$ contains $S_{j,\ell}$.
    \end{lemma}
    \begin{proof}
        Suppose, for the sake of contradiction, that there exist indices $i$ and $\ell$ such that neither condition (i) nor condition (ii) holds. That is, $x_{i,\ell} \notin M$, and for every index $j$ such that $x_i \in S_j$, vertex $S_{j,\ell}$ is not contained in $M$. 
        We show that the edge $(x_{i,\ell}, y_i)$ is not monitored by $M$.
        Let $\{a,b\}$ be a pair of vertices in $M$ that monitors $(x_{i,\ell},y_i)$, and let $P$ be a shortest path between $a$ and $b$. Since $x_{i,\ell} \notin M$ by our assumption, $x_{i,\ell}$ must be an internal vertex of $P$.
        One of the neighbors of $x_{i,\ell}$ in $P$ is $y_i$ while the other is a vertex $S_{j,\ell}$ for some $j$, which implies $x_i \in S_j$. 
        Since $S_{j,\ell} \notin M$, vertex $S_{j,\ell}$ cannot be an endvertex of $P$. 
        Let $w$ be the neighbor of $S_{j, \ell}$ other than $x_{i, \ell}$ in $P$.
        By the suboptimality of shortest paths, $\{w, y_i\}$ monitors $(x_{i,\ell}, y_i)$.
        We now provide a path from $w$ to $y_i$ of length at most $3$ which avoids $(x_{i,\ell}, y_i)$, thus contradicting either the fact that $\pi$ is a shortest path, or the fact that $\{w, y_i\}$ monitors $(x_{i,\ell}, y_i)$.  We consider two cases depending on whether $w \in Z$.
        \begin{description}
            \item[Case 1: $w \in Z$.] In this case, we must have $w = z_j$. Let $\ell'$ be any integer in $\{1, \dots, k\} \setminus \{\ell\}$ (observe that such an $\ell'$ exists since $k \ge 2$). An alternative shortest path is given by $\langle  z_j, S_{j, \ell'}, x_{i, \ell'}, y_i \rangle$.
            \item[Case 2: $w\not\in Z$.] In this case, $w = x_{i', \ell}$ for some $i' \neq i$. An alternative shortest path is given by $\langle x_{i', \ell}, y_{i'}, y_*, y_i \rangle$. \qedhere
        \end{description}
    \end{proof}

    \begin{lemma}\label[lemma]{lemma:meg_set_contains_set_cover}
        Given a \megset{} $M'$ of $G$, we can compute in polynomial time a \megset{} $M$ of $G$ such that $|M|\leq |M'|$ and, for every $\ell=1,\ldots,k$, the set $\mathcal{S}_\ell = \{S_j \in \mathcal{S} \mid S_{j,\ell} \in M\}$ is a set cover of $\mathcal{I}$. 
    \end{lemma}
    \begin{proof}
        By \Cref{lemma:meg_set_monitors_set_cover}, for every $i=1,\ldots,\eta$ and every $\ell=1,\ldots,k$, $M'$ contains $x_{i,\ell}$ or some $S_{j,\ell}$ such that $S_j$ covers $x_i$. We compute $M$ from $M'$ by replacing each $x_{i,\ell} \in M'$ with $S_{j,\ell}$, where $S_j \in \mathcal{S}$ is any set that covers $x_i$. As a consequence, for every $\ell=1,\ldots, k$, the set $\mathcal{S}_\ell = \{S_j \in \mathcal{S} \mid S_{j,\ell} \in M\}$ is a set cover of $\mathcal{I}$.
        Moreover, since $M'$ contains all vertices in $L$ by \Cref{lemma:leaf}, so does $M$. Then, \Cref{lemma:set_cover_induces_meg_set} implies that $M$ is a \megset{} of $G$.
    \end{proof}

    \begin{lemma}
    \label{lemma:inapx}
    Let $\alpha(n) \ge 1$ be a non-decreasing function such that $\alpha(n) \le \log n$, and let $\varepsilon$ be a positive constant of choice.
    Any polynomial-time $\alpha(n)$-approximation algorithm for the minimum \megset{} problem implies the existence of a  polynomial-time $(\alpha(\eta^7) + \varepsilon)$-approximation algorithm for \textsc{Set Cover with intersection $1$}.
    \end{lemma}
    \begin{proof}
        Consider an instance $\mathcal{I}=\langle X, \mathcal{S}\rangle$ of \textsc{Set Cover with intersection $1$} where $|X|=\eta$ and $|\mathcal{S}| = h \le \eta^2$.
    
        Let $c$ be a constant such that $7 \log x \le \varepsilon x$ for all $x \ge c$. W.l.o.g., we assume that $\eta \ge \max\{9, n_0\}$, since otherwise an optimal set cover can be found by brute force.
        Let $h^*$ be the size of an optimal set cover of $\mathcal{I}$. Recall that $N = \eta + h \le 2 \eta^2$.

        We now construct the graph $G$ with $n = (k+2) N + 2 \le \eta^7$ vertices by making $k = N(N+1)$ copies of $H$.
        Next, we run the $\alpha(n)$-approximation algorithm to compute a \megset{} $M'$ of $G$, and we use \Cref{lemma:meg_set_contains_set_cover} to find a \megset{} $M$ with $|M| \le |M'|$ that induces $k$ set covers $\mathcal{S}_1,\ldots,\mathcal{S}_k$ in polynomial time. Among these $k$ set covers, we output one $\mathcal{S}'$ of minimum size. 
        To analyze the approximation ratio of the above algorithm, let $M^*$ be an optimal \megset{} of $G$. 
    \Cref{lemma:set_cover_induces_meg_set} ensures that $|M^*| \leq |L| + kh^* = N + 1 + kh^* = k(h^* + \frac{1}{N})$, and hence:
    \begin{equation*}
        |M| \le |M'| \le \alpha(n) \cdot |M^*| \le \alpha(n) \cdot k \left(h^* + \frac{1}{N}\right). 
    \end{equation*}
    \noindent Then, using $\alpha(\eta^7) \le \log \eta^7 = 7 \log \eta$, the size of the returned set cover $\mathcal{S}'$ is:
    \begin{align*}
        |\mathcal{S}'| &\le \frac{|M|}{k} \le \alpha(n)  \cdot \left(h^* + \frac{1}{N}\right) \le 
        \alpha(\eta^7) \cdot h^* + \frac{\alpha(\eta^7)}{\eta} \le 
        \alpha(\eta^7) \cdot h^* + \frac{7 \log \eta}{\eta} \\ &\le 
        \alpha(\eta^7) \cdot h^* + \varepsilon
        \le (\alpha(\eta^7) + \varepsilon)h^*,
    \end{align*}
    hence $\mathcal{S}'$ is a $(\alpha(\eta^7) + \varepsilon)$-approximate solution for $\mathcal{I}$.
    \end{proof}

Since \cite{KumarAR00} showed that all polynomial-time (possibly randomized) approximation algorithms for \textsc{Set Cover with intersection $1$} have an approximation ratio of $\Omega(\log \eta)$, unless $\mathsf{NP} \subseteq \mathsf{ZTIME}(\eta^{O(\log \log \eta)})$, the Lemma~\ref{lemma:inapx} immediately implies the following inapproximability result for \megset:
\begin{theorem}
\label[theorem]{inapproximability_theorem}
All polynomial-time (possibly randomized) approximation algorithms for \megset must have an approximation factor of $\Omega(\log n)$, unless $\np \subseteq \mathsf{ZTIME}(n^{O(\log \log n)})$.
\end{theorem}

Observe that any \textsc{Vertex Cover} instance can be seen as an instance of \textsc{Set Cover with intersection $1$} in which edges are the items to be covered, and each vertex covers all its incident edges. Since each item (i.e., edge) is contained in exactly $2$ sets, we have that the degree of each vertex $x_{i,\ell}$ in the instance of \megset resulting from our reduction is exactly $3$. Moreover, once all such vertices are deleted, the resulting graph $G$ is acyclic which shows that $G$ has degeneracy at most $3$.

Since it is known that \textsc{Vertex Cover} cannot be approximated within any constant factor smaller than $10\sqrt{5} - 21 \approx 1.360679$, unless $\p = \np$, nor to within any constant factor smaller than $2$ if the unique games conjecture holds \cite{Khot02a, KhotR08}, Lemma~\ref{lemma:inapx} implies the following.

\begin{theorem}
    No polynomial-time approximation algorithm for \megset on graphs with degeneracy at most $3$ can have a constant approximation ratio smaller than $10\sqrt{5} - 21$, unless $\p=\np$.
    Moreover, no such algorithm can have a constant approximation ratio smaller than $2$, unless the unique games conjecture fails.
\end{theorem}

\section{Approximation algorithms}
\label{sec:apx_algs}

This section is devoted to designing our approximation algorithms.
It is convenient to work on a generalization of the \megset problem, in which each edge $e$ of $G$ has an associated non-negative \emph{weight} $w(e)$ and shortest paths are computed w.r.t.\ these weights. Moreover, each vertex $v$ of $G$ has a binary cost $c(v) \in \{0, 1\}$, and the input additionally specifies a subset $E'$ of the edges in $E$.
The goal is to find a set of vertices $M$ that monitors all edges in $E'$ and has minimum cost, defined as  $c(M) = \sum_{v \in M}{c(v)}$. We name this generalization \gmegset.

Notice that the case in which all vertex costs are $1$,  all edges weights are $1$, and $E' = E$ matches the standard definition of \megset, hence any approximation algorithm for \gmegset (whose approximation ratio is computed w.r.t.\ the cost of the returned set) also provides an approximation algorithm for \megset.

One can observe that a $\gmegset$ exists if and only if, for every edge $ (u,v) \in E'$, the shortest path between $u$ and $v$ is unique and consists of the sole edge $(u,v)$. Indeed, if this condition is met, the set $V$ is a trivial solution; otherwise, no subset of vertices can monitor edge $(u,v)$. As a consequence, in the rest of this section, we restrict ourselves to instances of \gmegset that satisfy the above property, which can be checked in polynomial time.

In the rest of this section, we first show a simple generalization of the $O(\sqrt{n \log n})$-approximation algorithm for \megset  \cite{Colli2023, BiloCF024, foucaud2024algorithmscomplexitymonitoringedgegeodetic} to \gmegset, and then we use it as a building block to develop a more involved algorithm that provides good approximate solutions for \gmegset on hereditary graph classes that admit small balanced vertex-separators.

\subsection{A simple approximation algorithm for \gmegset}
\label[algorithm]{simple_approximation_algorithm}

We start with the description of an $(2 c(M^*) \ln n)$-approximation algorithm for \gmegset, where $c(M^*)$ denotes the cost of an optimal \gmegset, that will be useful in the sequel. 

For technical convenience, we assume that all optimal solutions to the \gmegset instance select at least two vertices with non-zero cost.\footnote{Otherwise, there is at most one vertex with non-zero cost in an optimal solution, and such a vertex can be guessed in polynomial time.}

Let $Z$ be the set of all vertices of cost $0$. Observing that it is possible to check in polynomial time whether an edge of $E'$ is monitored by $Z$, we construct an instance of \textsc{Set Cover} in which the items are exactly the edges in $E'$ that are not monitored by $Z$.
The instance contains a set $S_p$ for each (unordered) pair $p = \{u,v\}$ of distinct vertices $u,v$ in $V \setminus Z$.
The set $S_p$ contains all items $e$ that are monitored by $Z \cup p$ (notice that an item in $S_p$ is not necessarily monitored by $p$). 

Next, we compute a $(\ln |E'|)$-approximate solution $\mathcal{S}'$ for such an instance in polynomial time.
This can be done by solving instances up to a constant size by brute force, while larger instances can be handled using the well-known greedy algorithm for \textsc{Set Cover} \cite{greedy_set_cover}, which has an approximation ratio of $\ln \eta - \ln \ln \eta + O(1)$, where $\eta$ is the number of items \cite{Slavik97}. 
We return the \gmegset $M$ obtained by selecting all vertices in $Z$ together with all vertices involved in at least one pair $p$ such that $S_{p} \in \mathcal{S}'$, i.e., $M = Z \cup \bigcup_{S_{p} \in \mathcal{S}'} p$.

Observe that, if $M^*$ is an optimal \gmegset, then the collection containing all sets $S_{p}$ for all unordered pairs $p$ of vertices in $M^* \setminus Z$ is a feasible solution $\mathcal{S}''$ to the \textsc{Set Cover} instance. Since $c(v) \in \{0,1\}$ for each $v \in V(G)$, if $\mathcal{S}^*$ is a minimum-cost set cover, we have  $|\mathcal{S}^*| \le \binom{|M^* \setminus Z|}{2} < \frac{|M^* \setminus Z|^2}{2}  = \frac{c(M^*)^2}{2}$.
Then:
\[
    c(M) \le 2 \cdot | \mathcal{S}' | \le  2 \cdot | \mathcal{S}^* |  \cdot \ln |E'|
    < c(M^*)^2 \ln |E'|
    \le 2 c(M^*)^2 \ln n. 
\]
We have proved the following lemma.
\begin{lemma}\label[lemma]{lm:initial_apx}
There exists a polynomial-time $(2 c(M^*) \ln n)$-approximation algorithm for \gmegset.
\end{lemma}

By observing that a trivial upper bound on the approximation ratio of $M$ is $\frac{c(V)}{c(M^*)}$, we have that the overall approximation ratio is upper bounded by
$
    \min\left\{2 c(M^*) \ln n,  \frac{c(V)}{c(M^*)} \right\} \le \sqrt{2 c(V) \ln n} ,
$
since the above minimum is maximized when $c(M^*) = \sqrt{ \frac{c(V)}{2 \ln n} }$.

\begin{lemma}\label[lemma]{lm:simple_apx}
There exists a polynomial-time $\sqrt{2 c(V) \ln n}$-approximation algorithm for \gmegset.
\end{lemma}
\Cref{lm:simple_apx} implies that $\megset$, in which $c(V)=n$, is $O(\sqrt{n \log n})$-approximable in polynomial time. 

\subsection{An approximation algorithm based on balanced graph separators}
\label[algorithm]{main_approximation_algorithm}

In this section, we present our main positive result. We start by giving some preliminary definitions and proving some useful structural properties of \gmegset{}s.

\subsubsection{Some structural properties}

Fix some instance of \gmegset with graph $G$, vertex costs $c(\cdot)$, edge weights $w(\cdot)$, and set of edges to be monitored $E'$.
We refer to a subset $C$ of vertices of $G$ as a \emph{cluster}, and we define the \emph{boundary} $\delta C$ of $C$ as $N_G(C) \setminus C$.
A subset $S$ of vertices is a periphery of $C$ if $\delta C \subseteq S \subseteq V \setminus C$, i.e., if $S$ contains all vertices in the boundary of $C$ but no vertex in $C$.
Given a pair $P = (C, S)$ consisting of a cluster $C$ and a periphery $S$ of $C$, we say that a path $\pi$ between two vertices $u, v$ in $G$ is a \emph{$P$-bypass} if all the following conditions are satisfied: (i) $u,v \in S$, (ii) $\pi$ is a shortest path in $G$, (iii) $\pi$ consists of at least two edges, and (iv) all internal vertices of $\pi$ belong to $V \setminus (C \cup S)$.
Notice that the existence of a $P$-bypass between two vertices $u, v$ in $G$ implies that $(u,v) \not \in E'$. 

\begin{figure}[t]
    \centering
    \includegraphics[scale=1]{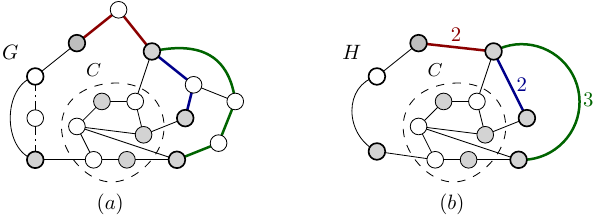}
    \caption{(a) The graph $G$ of an instance of \gmegset in which all edges have to be monitored, and a projection pair $P=(C,S)$ where the vertices in $S$ are shown in bold. Three $P$-bypasses are highlighted in red, blue, and green, respectively. All edge weights are unitary and are not shown. (b) The projection $H$ of $P$. There are three projection edges, highlighted in red, blue, and green, corresponding to the $P$-bypass shown with the same color in (a). The number next to a projection edge represents its weight, while non-projection edges have unit weight (not shown). Gray vertices form a \gmegset of $H$. The vertices in the periphery $S$ are shown in bold and have a cost of $0$. Notice how the gray vertices monitor the set $E_P$ of all non-projection edges of $H$ in $G$, and that there are some edges in $E \setminus E_P$ which are not monitored in $G$ (shown with dashed lines).}
    \label{fig:clusters}
\end{figure}

We refer to such a pair $P=(C,S)$ as a \emph{projection pair}, and we define the \emph{projection} of $P$ as the following instance of \gmegset.
The instance's graph $H$ has a vertex set consisting of all vertices in $C \cup S$, where vertices in $C$ retain their original cost in $G$, while vertices in $S$ have a cost of $0$. 
The edge set of $H$ consists of all edges in $G[ C \cup S ]$ with their original weights in $G$, plus some additional \emph{projection edges}.
More precisely, for each pair of distinct vertices $u,v \in S$ with at least one $P$-bypass between them, we add a \emph{projection edge} $e = (u,v)$ of weight $d_G(u,v)$, and we associate $e$ with an arbitrarily selected $P$-bypass $b_e$ between $u$ and $v$.
We refer to $H$ as the \emph{projection graph} of $P$, and we use $E_P$ to denote the set of all non-projection edges of $H$.
The edges of $H$ to be monitored are all those in $E_P \cap E'$. See \Cref{fig:clusters} for an example.

We provide different structural properties of walks in $G$ and the projection graph $H$ of a projection pair $P=(C,S)$. To distinguish between the weights of such walks, we will add the graph as a subscript to the weight function. 

\begin{lemma}
    \label[lemma]{lemma:mapping_walk}
    Let $P=(C,S)$ be a projection pair, and let $H$ be the projection graph of $P$. The following claims hold for any $u,v \in C \cup S$.
    \begin{enumerate}
        \item[(a)] For every walk $\pi$ between $u$ and $v$ in $H$, there exists a walk $\pi'$ between $u$ and $v$ in $G$ such that $w_H(\pi) = w_G(\pi')$ and $E(\pi) \cap E_P = E(\pi') \cap E_P$;
        \item[(b)] For every shortest path $\pi$ between $u$ and $v$ in $G$, there exists a walk $\pi'$ between $u$ and $v$ in $H$ such that $w_G(\pi) = w_H(\pi')$ and $E(\pi) \cap E_P = E(\pi') \cap E_P$.
    \end{enumerate}
\end{lemma}
\begin{proof}
    We start by proving (a). Given  a walk $\pi$ between $u$ and $v$ in $H$, we build the walk $\pi'$ 
    from $u$ to $v$ in $G$ from $\pi$ by replacing each projection edge $e \in \pi$ with its associated $P$-bypass $b_e$. 
    By construction, $E(\pi) \cap E_P = E(\pi') \cap E_P$.
    Moreover, for each edge $e \in E_P$, $w_H(e)=w_G(e)$  by construction, while for each projection edge $e$, $w_H(e) = w_G(b_e)$. Therefore,  $w_H(\pi) = w_G(\pi')$.
    
    We now prove (b). Given a shortest path $\pi$ between $u$ and $v$ in $G$, we decompose $\pi$ into an alternating sequence of walks $\pi_1, \widehat{\pi}_1, \pi_2, \widehat{\pi}_2, \dots, \pi_k$, where $\pi_i$ is a maximal subpath of $\pi$ that contains only edges in $E_P$, and $\widehat{\pi}_i$ is a maximal subpath of $\pi$ that contains only edges not in $E_P$.
    Let $u_i, v_i$ be the endvertices of $\widehat{\pi}_i$. We build a walk $\pi'$ from $u_i$ to $v_i$ in $H$ from $\pi$ by replacing each $\widehat{\pi}_i$ with the projection edge $e_i = (u_i, v_i)$ associated with a $P$-bypass $b_{e_i}$ between $u_i$ and $v_i$. By construction,  $E(\pi) \cap E_P = E(\pi') \cap E_P$. Moreover, for each edge $e \in E_P$, $w_H(e)=w_G(e)$  by construction. Now, since $\pi$ is a shortest path from $u$ to $v$ in $G$,  $\widehat{\pi}_i$ is a $P$-bypass in $G$, which implies $w_H(e_i)=w_G(\widehat{\pi}_i)$. Therefore, $w_G(\pi) = w_H(\pi')$.
\end{proof}

\begin{corollary}
    \label[corollary]{cor:mapping_sp}
    Let $P=(C, S)$ be a projection pair, and let $H$ be the projection graph of $P$. The following claims hold for any $u,v \in C \cup S$.
    \begin{enumerate}
        \item[(a)] For every shortest path $\pi$ between $u$ and $v$ in $H$, there exists a shortest path $\pi'$ between $u$ and $v$ in $G$ such that $w_H(\pi) = w_G(\pi')$ and $E(\pi) \cap E_P = E(\pi') \cap E_P$;
        \item[(b)] For every shortest path $\pi$ between $u$ and $v$ in $G$, there exists a shortest path $\pi'$ between $u$ and $v$ in $H$ such that $w_G(\pi) = w_H(\pi')$ and $E(\pi) \cap E_P = E(\pi') \cap E_P$.
    \end{enumerate}
\end{corollary}
\begin{proof}
    We start by proving (a). Let $\pi$ be a shortest path between $u$ and $v$ in $H$. 
    By \Cref{lemma:mapping_walk}~(a), there exists a walk $\pi'$ between $u$ and $v$ in $G$ such that  $w_H(\pi) = w_G(\pi')$  and $E(\pi) \cap E_P = E(\pi') \cap E_P$.
    We now argue that $\pi'$ is actually a shortest path in $G$.
    Suppose, towards a contradiction, that there exists some path $\pi''$ between $u$ and $v$ in $G$ such that $w_G(\pi'') < w_G(\pi')$.
    By \Cref{lemma:mapping_walk}~(b), this implies the existence of a walk $\pi'''$ between $u$ and $v$ in $H$ such that $w_H(\pi''') = w_G(\pi'')$, yielding the contradiction $w_H(\pi) = w_G(\pi') > w_G(\pi'') = w_H(\pi''')$.

    We now prove (b). Let $\pi$ be a shortest path between $u$ and $v$ in $G$. 
    By \Cref{lemma:mapping_walk}~(b), there exists a walk $\pi'$ between $u$ and $v$ in $H$ such that  $w_G(\pi) =w_H(\pi')$ and $E(\pi) \cap E_P = E(\pi') \cap E_P$.
    We now argue that $\pi'$ is actually a shortest path in $H$.
    Suppose, towards a contradiction, that there exists some path $\pi''$ between $u$ and $v$ in $H$ such that $w_H(\pi'') < w_H(\pi')$.
    By \Cref{lemma:mapping_walk}~(a), this implies the existence of a walk $\pi'''$ between $u$ and $v$ in $G$ such that $w_G(\pi''') = w_H(\pi'')$, yielding the contradiction $w_G(\pi) = w_H(\pi') > w_H(\pi'') = w_G(\pi''')$.
\end{proof}

\begin{corollary}
    \label[corollary]{cor:monitored_pruned_iff}
    Let $P$ be a projection pair, and let $H$ be the projection graph of $P$. 
    Let $e \in E_P$ be a non-projection edge in $H$, and let $u,v$ be a pair of vertices in $H$.
    Edge $e$ is monitored by the pair $u,v$ in $G$ if and only if it is monitored by $u,v$ in $H$. 
\end{corollary}
\begin{proof}
    Assume that $e$ is monitored by the pair $u,v$ in $G$, and let $\pi$ be a shortest path between $u$ and $v$ in $G$.
    By \Cref{cor:mapping_sp}~(b), there exists a shortest path $\pi'$ between $u$ and $v$ in $H$ that traverses $e$.
    We now argue that \emph{all} shortest paths between $u$ and $v$ in $H$ contain $e$.
    Indeed, if $H$ contained a shortest path $\pi''$ between $u$ and $v$ avoiding $e$, then \Cref{cor:mapping_sp}~(a) would imply the existence of a shortest path between $u$ and $v$ avoiding $e$ in $G$, i.e., $u,v$ would not monitor $e$ in $G$.

    Assume now that $e$ is monitored by the pair $u,v$ in $H$, and let $\pi$ be a shortest path between $u$ and $v$ in $H$.
    By \Cref{cor:mapping_sp}~(a), there exists a shortest path $\pi'$ between $u$ and $v$ in $G$ that traverses $e$.
    We now argue that \emph{all} shortest paths between $u$ and $v$ in $G$ contain $e$.
    Indeed, if $G$ contained a shortest path $\pi''$ between $u$ and $v$ avoiding $e$, then \Cref{cor:mapping_sp}~(b) would imply the existence of a shortest path between $u$ and $v$ avoiding $e$ in $H$, i.e., $u,v$ would not monitor $e$ in $H$.
\end{proof}

\begin{lemma}
\label[lemma]{lemma:from_G_to_C}
Let $P = (C, S)$ be a projection pair, and let $M$ be a \gmegset of $G$. 
Then, $(M \cap C)  \cup S$ is a \gmegset for the projection of $P$.
\end{lemma}
\begin{proof}
    Consider any edge $e \in E_P$ that belongs to $E'$, let $u',v'$ be a pair of vertices in $M$ that monitors $e$ in $G$, and let $\pi$ be a shortest path from $u'$ to $v'$ in $G$.
    Let $u$ (resp.\ $v$) be the last vertex in $(M \cap C)  \cup S$ encountered during a traversal of the subpath $\pi_u$ (resp.\ $\pi_v$) of $\pi$ from $u'$ (resp.\ $v'$) to the closest endvertex of $e$.
    Observe that $u$ (resp.\ $v$) exists since either $u' \in C \cup S$, or $\pi_u$ (resp.\ $\pi_v$) has the endvertex $u'$ (resp.\ $v'$) that is not in $C \cup S$ and the other endvertex in $C \cup S$, hence it must traverse at least one vertex in $S$. 

    Since the subpath of $\pi$ from $u$ to $v$ contains edge $e$, the pair $u,v$ monitors $e$ in $G$.
    Hence, by \Cref{cor:monitored_pruned_iff}, $u,v$ monitors $e$ in the projection of $P$.
\end{proof}

\begin{lemma}
    \label[lemma]{lemma:from_C_to_G}
    Let $P$ be a projection pair.
    Any \gmegset for the projection of $P$ monitors all edges in $E_P \cap E'$ in $G$.
\end{lemma}
\begin{proof}
    Let $M$ be a \gmegset for the projection of $P$, and let $H$ be the corresponding graph.
    We consider a generic edge $e \in E_P \cap E'$ and we prove that $e$ is also monitored by $M$ in $G$.
    Let $u,v \in M$ be a pair of vertices that monitors $e$ in $H$. 
    By \Cref{cor:monitored_pruned_iff}, the pair $u,v$ monitors $e$ in $G$.
\end{proof}

\subsubsection{Algorithm description}

In this section, we describe our main approximation algorithm that, as we will show, computes better solutions than the $(\sqrt{2c(V) \ln n})$-approximation of \Cref{lm:simple_apx} for the class of hereditary graphs that admit small balanced vertex-separators. Before delving into the details of the algorithm, we provide some basic definitions.

\begin{definition}
Given $\alpha \in (0, \frac{1}{2})$, an \emph{$\alpha$-balanced vertex-separator} of a graph $G = (V, E)$ is a partition of $V$ into three non-empty sets $S, A, B$, such that (i) every path from a vertex in $A$ to a vertex in $B$ traverses at least one vertex in $S$, and (ii) $\min\{|A|, |B|\} \ge \alpha |V|$. The size of the separator is $|S|$.
\end{definition}

Given $\beta \ge 0$, we say that a class $\mathcal{G}$ of graphs has \emph{efficiently computable $\alpha$-balanced separators} of size $O(n^\beta)$ if (i) $\mathcal{G}$ is hereditary, i.e., given a graph $G \in \mathcal{G}$, all induced subgraphs of $G$ are in $\mathcal{G}$, (ii) there are constants $n_0, \alpha > 0$, $\beta_0 \ge 1$ such that any graph $G \in \mathcal{G}$ with $n \ge n_0$ vertices admits an $\alpha$-balanced vertex-separator of size at most $\beta_0 n^\beta$, and (iii) such a separator can be found in polynomial time. We say that a graph is $\alpha$-non-separable if it admits no $\alpha$-balanced vertex-separator.
Similarly, we say that a cluster $C$ of $G$ is $\alpha$-non-separable if $G[C]$ is $\alpha$-non-separable.
Notice that if a class of graphs $\mathcal{G}$ has efficiently computable $\alpha$-balanced separators, $G \in \mathcal{G}$, and $C$ is an $\alpha$-non-separable cluster of $G$, then $|C| = O(1)$.

We consider a class $\mathcal{G}$ with efficiently computable $\alpha$-balanced separators of size $\beta_0 | V(G) |^{\beta}$ for some constants $\beta_0 \ge 1$ and $\beta \ge 0$, and we restrict our attention (w.l.o.g.) to instances of \gmegset whose input graph $G=(V,E)$ belongs to $\mathcal{G}$ and satisfies $c(V) \ge 2$.\footnote{Otherwise, when $c(V) \le 1$, an optimal solution consists either of all vertices of cost $0$, or of all vertices of cost $0$ plus the sole vertex of cost $1$.} Our algorithm works in two phases.

\subparagraph*{First phase.} In the first phase, we compute a hierarchical decomposition of $G$ into projection pairs, which can be represented as a binary tree $T$. Let $Z$ be the set of all vertices of cost $0$ in the \gmegset instance.
The root $r$ of the tree corresponds to the projection pair $P = (V \setminus Z, Z)$ and intuitively represents the whole input instance, as the projection of $P$ is the instance itself.
Given any vertex $v$ of the tree having $P = (C_v, S_v)$ as its projection pair, we have that $v$ is a leaf in $T$ whenever $C_v$ is $\alpha$-non-separable or the projection $H_v$ of $P$ admits a solution of cost $0$.
Otherwise, $v$ has exactly two children associated with the projection pairs $(A, S_v \cup S'_v)$ and $(B, S_v \cup S'_v)$, where $S'_v, A, B$ is an $\alpha$-balanced vertex separator of $G[C_v]$.

\begin{figure}[t]
    \centering
    \includegraphics[scale=1.1]{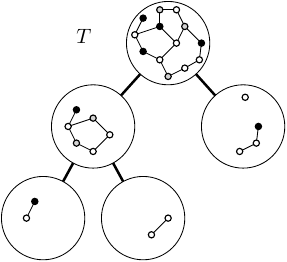}
    \caption{The tree $T$ representing a possible hierarchical decomposition of an input \gmegset instance in which all vertices have cost $1$. 
    Each vertex $v$ of $T$ contains the subgraph of $G$ induced by the corresponding cluster $C_v$, and the graph in the root node coincides with $G$.
    The parameters of the decomposition are $\beta_0 = 1$, $\beta=\frac{1}{2}$, $\alpha = \frac{1}{3}$.
    Vertices in the computed separators are shown in gray in the corresponding cluster. Black vertices in a leaf cluster $v$ of $T$ form an optimal solution to projection $H_v$ (not shown) when combined with the vertices in the corresponding periphery $S_v$ (having cost $0$).
    Observe that one of the instances admits a solution of cost $0$. Gray and black vertices in an internal cluster $v$ form a feasible \gmegset of the corresponding projection $H_v$ (not shown) when combined with those in $S_v$.}
    \label{fig:decomposition}
\end{figure}

\subparagraph*{Second phase.} In the second phase of our algorithm, we visit the tree in a bottom-up fashion and, for each examined vertex $v$ associated to the projection pair $P = (C_v, S_v)$, we compute a $\rho(h_v, n_v)$-approximate solution for the projection $H_v$ of $P$. Here, $\rho(h_v, n_v)$ is a function that depends on the number $n_v = |C_v|$ of vertices in cluster $C_v$ and, possibly, on the height $h_v$ of $v$ in $T$.\footnote{The height of vertex $v$ in $T$ is the height of the subtree of $T$ rooted in $v$, i.e., the maximum number of edges in the longest path from $v$ to a (not necessarily proper) descendant of $v$.}
Observe that all vertices in $C_v$ have cost $1$ w.r.t.\ the cost function $c_v$ of $H_v$, i.e., $n_v = c_v(C_v)$. 

In detail, if $v$ is a leaf in $T$, then either there exists an optimal \gmegset of $H_v$ with cost $0$, or $n_v= O(1)$, and we can find an optimal \gmegset of $H_v$ in constant time by brute force (possibly both).
Regardless of the case, we have an optimal solution for $H_v$, which is a $\rho(0, n_v)$-approximation for any $\rho(0, n_v) \ge 1$.

When the examined vertex $v$ of $T$ is a generic internal vertex at height $h_v$ associated to the projection pair $(C_v, S_v)$ and having $a$ and $b$ as children, we compute a \gmegset $M_v$ for $H_v$ by choosing the set of smallest cost between (i) a \gmegset $M'_v$ obtained as the union of $S_v$, a $\rho(h_v-1, n_a)$-approximate \gmegset $M_a$ for $H_a$, and a $\rho(h_v-1, n_b)$-approximate \gmegset $M_b$ for $H_b$, and (ii) a $(2 c_u(M_v^*) \ln |V(H_v)|)$-approximate solution computed using the algorithm of \Cref{lm:initial_apx}, where $M_v^*$ denotes an optimal solution for $H_v$ and $c_v$ is the cost function of $H_v$.
The next lemma provides an upper bound on the cost of $c_v(M'_v)$.

\begin{lemma}
    \label[lemma]{lemma:cost_M_u}
    Let $v$ be an internal vertex of $T$ having height $h_v$ and children $a$ and $b$. If $\rho(h_v-1, x)$ is monotonically non-decreasing w.r.t.\ $x$, 
    $M_a$ is a $\rho(h_v-1, n_a)$-approximate \gmegset for $H_a$, and $M_b$ is a $\rho(h_v-1, n_b)$-approximate \gmegset for $H_b$,
    then $M'_v$ is a \gmegset for $H_v$ with $c_v(M'_v) \le |S'_v| + c_v(M^*_v) \rho( h_v-1, (1-\alpha) n_v )$, where $M^*_v$ is an optimal solution for $H_v$.
\end{lemma}
\begin{proof}
    Let $M^*_a = (M^*_v \cap C_a) \cup S_a$ (resp.\ $M^*_b = (M^*_v \cap C_b) \cup S_b$).
    By \Cref{lemma:from_G_to_C}, $M^*_a$ and $M^*_b$ are \gmegset for $H_a$ and $H_b$, respectively.
    Then, $c_a( M^*_a ) = c_a(M^*_v \cap C_a) + c_a(S_a) = c_a(M^*_v \cap C_a) = c_v(M^*_v \cap C_a)$, and a similar derivation shows that $c_b( M^*_b ) = c_v(M^*_v \cap C_b)$.

    Since $M_a$ and $M_b$ are a $\rho(h_v-1, n_a)$- and a $\rho(h_v-1, n_b)$-approximation of $H_a$ and $H_b$, respectively, we have
    $c_a(M_a) \le c_v(M^*_v \cap C_a) \rho(h_v-1, n_a)$ and $c_b(M_b) \le c_v(M^*_v \cap C_b) \rho(h_v-1, n_b)$.

    Let $E'_v$ be the set of edges to be monitored in $H_v$. By \Cref{lemma:from_C_to_G}, $M_a$ monitors all edges of $E'_v$ in $H_v[C_a \cup S_a]$, and $M_b$ monitors all edges of $E'_v$ in  $H_v[C_b \cup S_b]$; therefore, $M'_v = M_a \cup M_b \cup S_v$ is a \gmegset for $H_v$. Then: 
    \begin{align*}
        c_v(M'_v) &\le c_v(M_a \cup M_b \cup S_v) = c_v(M_a \cup M_b)  
        \\& \le c_v(S'_v)  + c_v(M_a \setminus S'_v) + c_v(M_b \setminus S'_v) \\
        &\le |S'_v| + c_a(M_a) + c_b(M_b) \\
        &\le |S'_v| + c_v( M^*_a \cap C_a ) \rho(h_v-1, n_a) + c_v( M^*_b \cap C_b ) \rho(h_v-1, n_b) \\
        &\le |S'_v| + \big( c_v( M^*_a \cap C_a ) + c_v( M^*_b \cap C_b )  \big) \rho(h_v-1, (1-\alpha)n_v) \\
        &\le |S'_v| + c_v(M^*)  \rho(h_v-1, (1-\alpha)n_v). \qedhere
    \end{align*}
\end{proof}

The following two lemmas upper bound the approximation ratio achieved by $M_v$ for the instance $H_v$.
\begin{lemma}
    \label[lemma]{lemma:rho_general_beta_gt_0}
    Let $\rho(h, x) = x^{\beta/2} \cdot \frac{\sqrt{2 \beta_0 \ln n}}{1 - (1-\alpha)^{\beta / 2}}$.
    If $\beta > 0$, then $M_v$ is a $\rho(h_v, n_v)$-approximate \gmegset for $H_v$.
\end{lemma}
\begin{proof}
    The proof is by induction on $h_v$.
    The claim is trivially true for $h_v=0$ since, for a leaf $v$ of the tree $T$, we always compute an optimal \gmegset of $H_v$, and $\rho(0, n_v) = n_v^{\beta/2}\cdot\frac{\sqrt{2\beta_0 \ln n}}{1 - (1-\alpha)^{\beta/2}} \ge \sqrt{2 \ln 2} \ge 1$. Therefore, we assume that the claim holds for $h_v-1 \ge 0$ and we prove that it holds for $h_v$. 
    Consider a vertex $v$ at height $h_v$ in $T$ having $a$ and $b$ as its two children. 
    Let $M^*_v$ be an optimal \gmegset for $H_v$. By \Cref{lemma:cost_M_u}, we have: 
    \begin{align*}
        c_v(M'_v) &\le  |S'_v| + c_v(M^*_v) \rho( h_v-1, (1-\alpha) n_v ) \\
        &\le \beta_0 n_v^\beta + c_v(M_v^*)  (1-\alpha)^{\beta/2} n_v^{\beta/2} \cdot \frac{\sqrt{2 \beta_0 \ln n}}{1 - (1-\alpha)^{\beta / 2}}. 
    \end{align*}

    Since the returned solution is the one of minimum cost between $M'_v$ and a solution with approximation ratio 
    $2 c(M_v^*) \ln |V(H_u)| \le 2 c(M_v^*) \ln n$, the resulting approximation ratio is:
    \begin{equation*}
    \begin{aligned}
        \min\bigg\{
            & \frac{\beta_0 n_v^\beta}{c_v(M^*_v)} +  (1-\alpha)^{\beta/2} n_v^{\beta/2}\cdot\frac{\sqrt{2 \beta_0 \ln n}}{1 - (1-\alpha)^{\beta / 2}}, 2 c_v(M^*_v) \ln n
        \bigg\} \\
        &\le 
        \min\left\{
            \frac{\beta_0 n_v^\beta}{c_v(M^*_v)}, 2 c_v(M^*_v) \ln n
        \right\} + (1-\alpha)^{\beta/2} n_v^{\beta/2} \cdot \frac{\sqrt{2 \beta_0 \ln n}}{1 - (1-\alpha)^{\beta / 2}} \\
         &\le \sqrt{2\beta_0 \ln n} \cdot n_v^{\beta/2} + (1-\alpha)^{\beta/2} n_v^{\beta/2} \frac{\sqrt{2 \beta_0 \ln n}}{1 - (1-\alpha)^{\beta / 2}} \\ 
        & = \sqrt{2 \beta_0 \ln n} \cdot n_v^{\beta/2} \cdot \left(1 + \frac{(1-\alpha)^{\beta/2}}{1-(1-\alpha)^{\beta/2}} \right)
        = n_v^{\beta/2} \cdot \frac{\sqrt{2 \beta_0 \ln n}}{1-(1-\alpha)^{\beta/2}},
    \end{aligned}
    \end{equation*}
    where, in the second inequality, we used the fact that $\min\left\{
            \frac{\beta_0 n_v^\beta}{c_v(M^*_v)}, 2 c_v(M^*_v) \ln n\right\}$ is maximized for $\frac{\beta_0 n_v^\beta}{c_v(M^*_v)}=2 c_v(M^*_v) \ln n$, i.e., when $c_v(M^*_v)=\sqrt{\frac{\beta_0 n_v^\beta}{2\ln n}}$.
\end{proof}

\begin{lemma}
    \label[lemma]{lemma:rho_general_beta_0}
    Let $\rho(h, x) = (h+1) \sqrt{2 \beta_0 \ln n}$.
    If $\beta = 0$, then $M_v$ is a $\rho(h_v, n_v)$-approximate \gmegset for $H_v$.
\end{lemma}
\begin{proof}
    The proof is by induction on $h_v$.
    The claim is trivially true for $h_v=0$ since, for a leaf $v$ of the tree $T$, we always compute an optimal \gmegset of $H_v$, $\rho(0, n_v) = \sqrt{n_u \beta_0 \ln n} \ge \sqrt{2 \ln 2} \ge 1$.
    
    Therefore, we assume that the claim holds for $h_v-1 \ge 0$ and we prove that it holds for $h_v$. 
    Consider a vertex $v$ at height $h_v$, let $a$ and $b$ be its two children, and let $S'_v$ be the corresponding separator of $G[C_v]$.
    Let $M^*_v$ be an optimal solution for $H_v$. By \Cref{lemma:cost_M_u}, we have: 
    \begin{align*}
        c_v(M'_v) \le  |S'_v| + c(M_v^*) \rho( h_v-1, (1-\alpha) n_v )
        \le \beta_0 n_v^\beta + c(M_v^*)  h_v \sqrt{2 \beta_0 \ln n}. 
    \end{align*}
    
    Since the returned solution is the one of minimum cost between $M'_v$ and a solution with approximation ratio 
    $2 c_v(M^*_v)  \ln |V(H_v)| \le 2 c_v(M^*_v) \ln n$, the resulting approximation ratio is:
    \begin{equation*}
    \begin{aligned}
        \min\bigg\{
            \frac{\beta_0}{c_v(M^*_v)}  + h_v & \sqrt{2 \beta_0 \ln n},  2 c_v(M^*_v)  \ln n
        \bigg\} \\
        &\le 
        \min\left\{
            \frac{\beta_0}{c_v(M^*_v)}, 2 c_v(M^*_v)  \ln n
        \right\} + h_v \sqrt{2 \beta_0 \ln n} \\
        &\le \sqrt{2\beta_0 \ln n}  + h_v \sqrt{2 \beta_0 \ln n} 
        = (h_v+1) \sqrt{2\beta_0 \ln n},
    \end{aligned}
    \end{equation*}
    where we used the fact that $\min\left\{ \frac{\beta_0}{c_v(M_v^*)}, 2 c_v(M_v^*)  \ln n \right\}$ is maximized for $c_v(M_v^*) =  \sqrt{\frac{\beta_0}{2 \ln n}}$. 
\end{proof}

\noindent We can finally state the main results of this section.
\begin{theorem}
    There exists a polynomial-time approximation algorithm for \gmegset with an approximation ratio of
    $O(c(V)^{\beta/2} \sqrt{\log n})$ 
    if $\beta > 0$ and an approximation ratio of 
    $O(\log c(V) \cdot \sqrt{\log n})$ 
    if $\beta = 0$.
\end{theorem}
\begin{proof}
    Recall that $r$ is the root of $T$.
    We argue that $M_r$ achieves the claimed approximation ratios for the projection $H_r$, which coincides with the input \gmegset instance since $C_r \cup S_r = V$.
    When $\beta > 0$, this follows directly from \Cref{lemma:rho_general_beta_gt_0} and from the fact that $G[C_r]$ has $n_r = c(V)$ vertices.
    Regarding the case $\beta = 0$, if $v$ is the parent of $u$, then $n_u \le (1-\alpha) n_v$.
    Then, the tree $T$ has height at most
    $\left\lceil \log_{\frac{1}{1-\alpha}} n_r \right\rceil
    = \left\lceil \frac{\log c(V)}{\log \frac{1}{1-\alpha}} \right\rceil
    $, and \Cref{lemma:rho_general_beta_0} implies that the achieved approximation is at most
    $\left(1 + \left\lceil \frac{\log c(V)}{\log \frac{1}{1-\alpha}} \right\rceil \right) \cdot \sqrt{2 \beta_0 \ln n}$.
\end{proof}

The above theorem implies the following approximability results for \gmegset (and for \megset, where $c(V) = n$):
\begin{corollary}
\gmegset is approximable in polynomial time on:    
\begin{itemize}
    \item $K_h$-minor-free graphs, with an approximation ratio of $O(c(V)^{\frac{1}{4}} \sqrt{\log n})$, where $K_h$ is a complete graph on $h$ vertices and $h=O(1)$, using the balanced separators of size $O(\sqrt{n})$ shown in \cite{abs-2512-01587}.
    \item planar graphs, with an approximation ratio of $O(c(V)^{\frac{1}{4}} \sqrt{\log n})$ using the previous result with $h=5$ and the fact that planar graphs are $K_5$-minor-free. 
    \item graphs with bounded treewidth, with an approximation ratio of $O(\log c(V) \sqrt{\log n})$, using the separator induced by a centroid bag of the corresponding tree decomposition \cite[Sections~12.3~and~12.4]{diestel};
    \item graphs with constant genus, with an approximation ratio of $O(c(V)^{\frac{1}{4}} \sqrt{\log n})$ using the balanced separators of size $O(\sqrt{n})$ shown in \cite{LiptonT80}; 
    \item graphs with apex $k = O(n^{1/4})$, with an approximation ratio of $O(c(V)^{\frac{1}{4}} \sqrt{\log n})$, since it is possible to find, in polynomial time, a set $S$ of $O(k)$ vertices such that $G[V(G) \setminus S]$ is planar \cite{apex_separators}.
\end{itemize}
\end{corollary}

\bibliographystyle{elsarticle-num} 
\bibliography{references}
\end{document}